# Higgs mode in the *d*-wave superconductor $Bi_2Sr_2CaCu_2O_{8+x}$ driven by an intense terahertz pulse

Kota Katsumi[1], Naoto Tsuji[2], Yuki I. Hamada[1], Ryusuke Matsunaga[1,3], John Schneeloch[4], Ruidan D. Zhong[4], Genda. D. Gu[4], Hideo Aoki[1,5,6], Yann Gallais[1,7,8] and Ryo Shimano[1,8]

[1] Department of Physics, The University of Tokyo, Tokyo, 113-0033, Japan
[2] RIKEN Center for Emergent Matter Science (CEMS), Wako, 351-0198, Japan
[3] JST, PREST, Kawaguchi, 332-0012, Japan
[4] Brookhaven National Lab, Upton, NY, 11973, USA
[5] Department of Physics, ETH Zürich, 8093 Zürich, Switzerland
[6] National Institute of Advanced Industrial Science and Technology (AIST), Tsukuba, 305-8568, Japan
[7] MPQ CNRS, Université Paris Diderot, Bâtiment Condorcet, 75205 Paris Cedex 13, France
[8] Cryogenic Research Center, The University of Tokyo, Tokyo, 113-0032, Japan

## Abstract

We investigated the terahertz (THz)-pulse driven nonlinear response in the *d*-wave cuprate superconductor $Bi_2Sr_2CaCu_2O_{8+x}$ (Bi2212) using a THz pump near-infrared probe scheme in the time domain. We have observed an oscillatory behavior of the optical reflectivity that follows the THz electric field squared and is strongly enhanced below $T_c$. The corresponding third-order nonlinear effect exhibits both $A_{1g}$ and $B_{1g}$ symmetry components, which are decomposed from polarization-resolved measurements. Comparison with a BCS calculation of the nonlinear susceptibility indicates that the $A_{1g}$ component is associated with the Higgs mode of the *d*-wave order parameter.

In a superconductor the spontaneous breaking of $U(1)$ phase leads to two types of collective excitations of the order parameter. One is the Nambu-Goldstone mode which is pushed up to the plasma frequency due to the Coulomb interaction, while the other is the amplitude (Higgs) mode in a conventional *s*-wave superconductor [1,2]. Being chargeless and spinless, the Higgs mode in superconductors only weakly couples to external probes, and has thus remained elusive experimentally until recently. It has been initially identified in a Raman measurement in NbSe2, where the charge density wave (CDW) coexists with superconductivity and makes the mode Raman-active via its indirect coupling to the CDW order parameter [3–5]. Recently, the Higgs mode has been clearly observed in a more generic situation (without CDW) in an *s*-wave superconductor $Nb_xTi_{1-x}N$ (NbN) by ultrafast terahertz (THz) pump-THz probe spectroscopy [6]. The role of ultrashort THz-pump pulse is to provide a non-adiabatic quench of the order parameter by instantaneously creating a population of unpaired quasiparticles (QPs) around the superconducting (SC) gap energy that triggers Higgs oscillations in the time domain [7]. The Higgs dynamics of the SC order parameter has since been theoretically studied in a variety of contexts, ranging from multiband to unconventional superconductors [8–12]. Specifically, in a *d*-wave superconductor such as the cuprates with nodes in the gap function, the Higgs mode was theoretically shown to decay much faster than in the *s*-wave case because of the presence of low-energy QPs [9]. Besides, in many unconventional superconductors the coexistence with other electronic orders and/or competing interactions can significantly alter the Higgs-mode dynamics, and may lead to a rich assortment of collective modes [8,13–16]. Thus, it is imperative to explore how the Higgs mode behaves in unconventional superconductor.

In this context, nonlinear optical effects have recently kicked off an alternative way to probe the Higgs mode [17,18]. This was demonstrated in the conventional *s*-wave superconductor NbN, where, remarkably, a resonance between the Higgs mode and an intense THz field with a photon energy $\omega$ below the SC gap $2\Delta$ was shown to induce large third-harmonic generation (THG) with a resonance condition $2\omega = 2\Delta$ [17,18]. It has subsequently been pointed out that, in addition to the Higgs mode, charge density fluctuations (CDF) can also contribute to the THG signal at the same frequency [19]. Within the BCS mean-field approximation, the contribution of CDF to THG should be much larger than the Higgs-mode contribution. More recently, the contributions from the Higgs mode and CDF have been decomposed in NbN via polarization-resolved measurements. The decomposition, theoretically shown to hold even beyond the BCS approximation, has revealed that the Higgs mode actually gives a dominant contribution to the THG far exceeding the CDF contribution [20]. Physically, the dominance of the Higgs mode in THG can be attributed to dynamical effects in the pairing, such as the retardation in the phonon-mediated electron interaction that are neglected in the BCS approximation [21]. Given this situation for the conventional *s*-wave superconductors, what happens in *d*-wave superconductors next becomes of great interest.

In this Letter, we report an observation of the third-order nonlinear signal in a *d*-wave cuprate superconductor $Bi_2Sr_2CaCu_2O_{8+x}$ (Bi2212) from THz pump-optical reflectivity probe measurements over a wide range of carrier doping. The third-order nonlinear signal, akin to a THz Kerr effect, turns out to manifest itself as an oscillatory behavior of the optical reflectivity that follows the squared THz electric field (E-field) with strong enhancement below $T_c$. The THz Kerr signal is here further decomposed into $A_{1g}$ and $B_{1g}$ symmetry components from polarization-resolved measurements. We then show that a comparison with BCS calculations for both Higgs-mode and CDF contributions to each symmetry component strongly indicates that the observed $A_{1g}$ component arises from the coupling of the d-wave order parameter to the Higgs mode.

We have performed a THz pump-optical probe (TPOP) measurements, schematically illustrated in Fig. 1(a), on freshly cleaved optimally-doped (OP90, $T_c \approx 90$ K) as well as overdoped (OD78, OD66 and OD52, with $T_c \approx 78, 66, 52$ K, respectively) and underdoped (UD74 and UD58, with $T_c \approx 74, 58$ K, respectively) Bi2212 single crystals grown with the floating-zone method. The description of the THz pulse generation is given in Supplemental Material (SM) [22]. For the probe we used a near-infrared pulse at 800 nm which has been widely used as a sensitive probe for investigating the dynamics of the SC state in the cuprates [28–36]. The measurements were performed as a function of both the pump and probe polarization angles $\theta_{Pump}$, $\theta_{Probe}$ as defined in Fig. 1(b). As we shall show, the polarization dependence of the pump-probe signal is crucial in discriminating the Higgs-mode and CDF contributions. The central frequency component of the THz-pump E-field is ~ 0.6 THz = 2.4 meV, which is much smaller than the anti-nodal SC gap energy, $2\Delta_0 > 20$ meV, in Bi2212 for the present doping levels [37,38]. This THz pulse does not significantly deplete the SC state, as evidenced by the absence of any sign of pump-probe signal-saturation up to ~ 350 kV/cm (see SM Fig. S2).

Let us start with the result for sample OP90. The THz pulse-induced transient reflectivity change $\Delta R$ for $\theta_{Pump} = \theta_{Probe} = 0°$ is shown in Fig. 1(c) at various temperatures. At 30 K below $T_c$, an oscillatory behavior of $\Delta R/R$ that follows the squared THz-pump E-field $|E_{Pump}(t)|^2$ is clearly identified. This quasi-instantaneous oscillatory component is similar to the forced oscillation of the order parameter observed in a conventional *s*-wave superconductor NbN, which also follows $|E_{Pump}(t)|^2$ [6]. Accordingly, the maximum amplitude of $\Delta R/R$ is proportional to the square of the peak THz-pump E-field as shown in SM Fig. S2. In addition to the oscillatory component, $\Delta R/R$ has a positive decaying component that survives up to at least ~ 10 ps. At 100 K slightly above $T_c$, the signal consists of a much weaker oscillatory component and a decaying signal that switches sign after ~ 4 ps. At 300 K the decaying signal remains positive at all delays.

The amplitude of $\Delta R/R$ as a function of $\theta_{\text{Probe}}$ at a fixed delay $t = 2$ ps at which the oscillatory component is maximum is displayed in Fig. 1(d). The $\Delta R/R$ is essentially independent of the angle at 300 K and 100 K. At 30 K below $T_c$, however, it displays significant dependence on $\theta_{\text{Probe}}$, which follows a form $A + B\cos(2\theta_{\text{Probe}})$. By contrast the $\Delta R/R$ signal at $t = 4$ ps does not show any polarization dependence at 30 K. Similar results were obtained when the pump polarization angle $\theta_{\text{Pump}}$ is varied with a fixed $\theta_{\text{Probe}} = 0°$, demonstrating the symmetrical roles played by the pump and probe polarization angles in the observed signal (see SM Fig. S3(a)).

The pump E-field and polarization dependences of the oscillatory component are consistent with a THz Kerr effect where the strong THz E-field modulates the optical reflectivity in the near-infrared (800 nm) regime [39]. This process is described by a third-order nonlinear susceptibility $\chi^{(3)}(\omega; \omega, +\Omega, -\Omega)$ [40], where $\omega$ and $\Omega$ are the frequencies of the near-infrared pulse and THz-pump pulse, respectively. The THz pulse-induced reflectivity change $\Delta R/R$ can be expressed in terms of $\chi^{(3)}$ (for details see SM) as

$$\frac{\Delta R}{R}(E_i^{\text{Probe}}, E_j^{\text{Probe}}) \sim \frac{1}{R}\frac{\partial R}{\partial \varepsilon_1} \varepsilon_0 \text{Re}\chi_{ijkl}^{(3)} E_k^{\text{Pump}} E_l^{\text{Pump}}, \quad (1)$$

where $E_i$ denotes the $i$th component of the THz-pump or probe E-field and $\varepsilon_1$ is the real part of the dielectric constant. Assuming tetragonal symmetry for Bi2212, we can analyze the polarization dependence of $\chi^{(3)}(\theta_{\text{Pump}}, \theta_{\text{Probe}})$ in terms of the irreducible representations of $D_{4h}$ point group as

$$\chi^{(3)}(\theta_{\text{Pump}}, \theta_{\text{Probe}}) = \frac{1}{2}(\chi_{A_{1g}}^{(3)} + \chi_{B_{1g}}^{(3)}\cos2\theta_{\text{Pump}}\cos2\theta_{\text{Probe}} + \chi_{B_{2g}}^{(3)}\sin2\theta_{\text{Pump}}\sin2\theta_{\text{Probe}}), \quad (2)$$

where we have defined $\chi_{A_{1g}}^{(3)} = \chi_{xxxx}^{(3)} + \chi_{xxyy}^{(3)}$, $\chi_{B_{1g}}^{(3)} = \chi_{xxxx}^{(3)} - \chi_{xxyy}^{(3)}$ and $\chi_{B_{2g}}^{(3)} = \chi_{xyxy}^{(3)} + \chi_{xyyx}^{(3)}$.

For a given $\theta_{\text{Pump}}$, the $A_{1g}$ and $B_{1g}$ signals respectively correspond to the isotropic and $\cos2\theta_{\text{Probe}}$ components observed in Fig. 1(d), which can be extracted by adding or subtracting $\Delta R/R$ ($\theta_{\text{Probe}} = 0°$) and $\Delta R/R$ ($\theta_{\text{Probe}} = 90°$). As expected from Eq. (2) the extracted $A_{1g}$ signal is found to be independent of $\theta_{\text{Pump}}$, while the $B_{1g}$ signal follows $\cos2\theta_{\text{Pump}}$ (see Fig. S3(b)). On the other hand, no $B_{2g}$ signal, obtained by subtracting $\Delta R/R$ ($\theta_{\text{Probe}} = 45°$) from $\Delta R/R$ ($\theta_{\text{Probe}} = -45°$), is observed within the noise level ($10^{-5}$) (see Fig. SM S4). We stress that the $B_{1/2g}$ THz Kerr signals discussed here are not linked to any symmetry breaking order, but simply follow from the general properties of susceptibility tensors for $D_{4h}$ point group.

In Figs. 2(a) and 2(b) we present the temperature dependences of the $A_{1g}$ and $B_{1g}$ signals for OP90. For both symmetries the signal strongly evolves below $T_c$ in the interval (1-2 ps), corresponding to the oscillatory component discussed above. The decaying component at longer delays ($t > 4$ ps) is only observed in $A_{1g}$ symmetry, and displays a more complex temperature dependence. To obtain more detailed information on the temperature dependence of the symmetry-resolved components, we fitted the transient signals with $|E_{Pump}(t)|^2$ (for the oscillatory component or THz Kerr signal), an error function (decaying component) and a step function (offset component). In addition, $|E_{Pump}(t)|^2$ was convoluted with an exponential function to take account of a small delay (~ 200 fs) in the nonlinear response of the system [41] (see SM). The fitted result at 10 K is shown with the solid curves in Fig. 2(c).

Figure 2(d) summarizes the amplitudes of the different components of the $A_{1g}$ and $B_{1g}$ signals against temperature. The $A_{1g}$ and $B_{1g}$ oscillatory components sharply increase below $T_c$, indicating the onset of a new channel in the THz Kerr response upon entering the SC state. Contrary to the $A_{1g}$ oscillatory component which remains positive at all the temperatures, the decaying component switches sign twice, at $T_c$ and $T^*$, respectively. Here $T^*$ is within the range of the pseudogap (PG) temperature as determined by ARPES [42]. The decaying component also displays a sharp maximum slightly below $T_c$. The overall behavior of the decaying component, including the sign changes, is in good agreement with previous optical pump-optical probe (OPOP) measurements [32]. In these measurements the positive decaying component below $T_c$ and the negative decaying component above $T_c$ were assigned to QP relaxation in the SC and PG states, respectively.

Let us now turn to what happens when the doping level is varied. In all the samples, an oscillatory component of $\Delta R/R$ was found to be strongly enhanced below $T_c$ (with $T$-dependent temporal behavior shown in SM Fig. S6 for UD74 and OD78). The result for $T$ = 10 K is shown in Figs. 3(a)-3(c) for UD74, OP90 and OD78, respectively. The $A_{1g}$ SC decaying component in OD samples is negative (Fig. 3(c) for OD78 and SM Fig. S7 for OD52 and OD66), while those in UD74 and OP90 are positive. The sign change close to the optimal doping is also in good agreement with that of OPOP measurements in Bi2212 [31], but contrasts with the $A_{1g}$ oscillatory component which remains positive for all the samples studied. In addition, it is also apparent that the amplitudes of the $A_{1g}$ and $B_{1g}$ oscillatory components in the SC state strongly depend on doping. Since $\Delta R/R$ depends on $\partial R/\partial \varepsilon_1$ at 800 nm that in turn depends on doping, the evolution of the symmetry component of $\chi^{(3)}$ with doping can be best tracked in terms of the ratio of the $B_{1g}$ and the $A_{1g}$ oscillatory components as summarized in Fig. 3(d) at 10 K. While the $B_{1g}$ component is more than an order of magnitude weaker than the $A_{1g}$ component in UD samples, it continuously increases with doping but never exceeds the $A_{1g}$ component in the doping range studied, $p$ = 0.10 - 0.22.

We now compare the observations with theoretical expectations, focusing on the origin of the symmetry-dependent THz Kerr signal observed in the SC state. As in the case of the THG, both CDF and Higgs mode can contribute to $\chi^{(3)}$. In Fig. 4(a), we show the diagrams [(i)-(iv)] that represent the CDF contributions to $\chi^{(3)}(\omega; \omega, +\Omega, -\Omega)$. The diagram (iv) does not show characteristic temperature dependence, and is irrelevant to superconductivity. In the case of TPOP measurements, the probe frequency $\omega$ exceeds all the other relevant energy scales ($\Omega$, $2\Delta_0$, $T_c$, etc.), where the contributions of the diagrams (i) and (ii) are suppressed ($\sim 1/\omega^2$) as compared to that of (iii). Hence the contribution relevant to the present experiment essentially comes from the frequency-independent diagram (iii).

Based on the above consideration, we indicate in Table 1 the general behavior of the symmetry decompositions for the CDF and Higgs mode, respectively (see SM for details). While the CDF appears in all the symmetry channels, the Higgs mode selectively appears in the $A_{1g}$ symmetry. To quantify the magnitudes of the CDF and Higgs-mode contribution in different symmetries, we employ the single-band tight-binding model for Bi2212 to calculate the $\chi^{(3)}$ diagram (iii) numerically within the mean-field treatment (see SM). In the result for CDF, displayed in Fig. 4(b), we can see that all the contributions grow below $T_c$, and hence are correlated to superconductivity. Within the CDF, the $B_{1g}$ symmetry exhibits by far the largest contribution. This can be generally understood from the microscopic expression for $\chi^{(3)}$ (see SM): The terms involved in the $A_{1g}$ symmetry have both positive and negative contributions in the anti-nodal regions of the Fermi surface, which partially cancel with each other, while those in the $B_{1g}$ symmetry have positive contributions. The $B_{2g}$ component is proportional to the square of the subdominant second-neighbor hopping, and becomes smaller than the other symmetry components. Although the respective weight of the Higgs-mode and CDF contributions can depend on the level of approximations used in theoretical treatments [20], we expect the symmetry dependence of the CDF contribution to be more robust because it is essentially tied to the band structure as explained above.

As we have seen, the $A_{1g}$ oscillatory component is experimentally dominant in all the samples studied. This strongly implies that the $A_{1g}$ oscillatory component originates from the Higgs-mode contribution of the $d$-wave order parameter, while the $B_{1g}$ oscillatory component likely originates from CDF. The absence of the $B_{2g}$ CDF component in our measurement agrees with the mean-field result, in which it is about 17 times smaller than the $B_{1g}$ contribution. The above interpretation is also supported by a comparison with Raman results in Bi2212, which are commonly attributed to CDF [43]. First, the increase in the relative amplitude of the $B_{1g}$ component with doping is consistent with the strong increase in the pair-breaking peak intensity observed in $B_{1g}$ Raman spectra toward $p =$ 0.22 [38,44]. Second, in underdoped Bi2212 samples both $B_{1g}$ and $A_{1g}$ SC Raman responses vanish, leaving only a weak $B_{2g}$ Raman signature of the SC state [45,46]. It was interpreted as a consequence of the PG opening which strongly suppresses the CDF response coming from anti-nodal QPs, but leaves intact the nodal QP probed in $B_{2g}$ response [46]. This contrasts strongly with the dominance of the $A_{1g}$ oscillatory component observed here in UD samples in the THz Kerr signal, and further reinforces our assignment as arising from the $d$-wave Higgs mode.

Precise physical origin of the dominance of the Higgs-mode contribution to the THz Kerr effect remains an open problem. This may be a general property of nonlinear susceptibilities in the SC state at THz frequencies, since the same observation was deduced from the polarization dependence of the THz THG signal in the conventional *s*-wave NbN [20]. A recent dynamical mean field theory (DMFT) calculation has shown, as mentioned above, that the Higgs-mode contribution can actually exceed the CDF contribution if retardation effects are considered in strongly electron-phonon-coupled superconductors [20]. An interesting question then is whether this also holds for unconventional superconductors.

In conclusion, we have studied THz pulse-induced nonequilibrium dynamics in Bi2212 from the change in the optical reflectivity. We observed an oscillatory behavior of the optical reflectivity proportional to $|E_{\text{Pump}}(t)|^2$ which we assign to a nonlinear THz Kerr effect. The signal is strongly enhanced below $T_c$, and is decomposed into the $A_{1g}$ and $B_{1g}$ components. A theoretical calculation of the relevant third-order nonlinear susceptibility indicates that the $A_{1g}$ component corresponds to the Higgs mode, while the $B_{1g}$ component originates from CDF. We envisage that THz nonlinear techniques will be a promising avenue for studying collective excitations of unconventional SC, which can possibly probe their interplay with other orders.

We acknowledge Dirk Manske for fruitful discussions. This work was partially supported by JSPS KAKENHI Grant Nos. 15H02102, 16K17729, JP26247057, and JP25800175 and by the Photon Frontier Network Program from MEXT, Japan. H.A. is also supported by ImPACT Program of Council for Science, Technology and Innovation, Cabinet Office, Government of Japan (Grant No. 2015-PM12-05-01) from JST. Y.G. acknowledges financial support for the Japan Society for the Promotion of Science (JSPS). Work at Brookhaven was supported by the Office of Basic Energy Sciences (BES), Division of Materials Sciences and Engineering, U.S. Department of Energy (DOE), through Contract No. de-sc0012704. RDZ and JS were supported by the Center for Emergent Superconductivity, an Energy Frontier Research Center funded by BES.

**Figure captions**

**Figure 1** (Color online) (a) A geometry for the TPOP measurements. (b) A schematic illustration of the $CuO_2$ plane, on which the pump ($\theta_{Pump}$) and probe ($\theta_{Probe}$) polarization angles are defined relative to the Cu-O bond. (c) THz pulse-induced transient reflectivity change $\Delta R/R$ at $\theta_{Probe} = 0°$ as a function of delay time at typical temperatures for OP90. Top panel shows the waveform of the squared THz E-field. (d) Probe polarization dependence (circles) of the amplitude of $\Delta R/R$ at fixed delays at various temperatures for OP90 when $\theta_{Pump} = 0°$. Curves show the fitting with a form $A + B \cos(2\theta_{Probe})$.

**Figure 2** (Color online) For OP90, temperature dependences of the $A_{1g}$ (a) and $B_{1g}$ (b) components of $\Delta R/R$. Red dashed lines indicate $T_c$. (c) The $A_{1g}$ and $B_{1g}$ components against the delay time at 10 K with fitting curves. (d) Temperature dependences of the $A_{1g}$ decaying component (blue), the $A_{1g}$ oscillatory component (red), and the $B_{1g}$ oscillatory component (green).

**Figure 3** (Color online) (a)-(c) Temporal behavior of the $A_{1g}$ and $B_{1g}$ components as compared for UD74, OP90, and OD78 at 10 K. Red and blue curves represent the $A_{1g}$ oscillatory component and decaying component, respectively, while black lines are total fitting curves. (d) Doping dependence of the ratio of the amplitude of the $B_{1g}$ and the $A_{1g}$ oscillatory components at 10 K (red circles; left axis) with $T_c$ (blue; right axis) for all the samples studied. The hole concentration $p$ is determined from $T_c$ with Presland and Tallon's equation [47].

**Figure 4** (Color online) (a) The diagrams representing the CDF contributions to $\chi^{(3)}(\omega; \omega, +\Omega, -\Omega)$. (b) Theoretical result for the CDF contribution for different symmetries.

**Table 1** General polarization dependence of CDF and Higgs-mode contributions for the TPOP measurements.

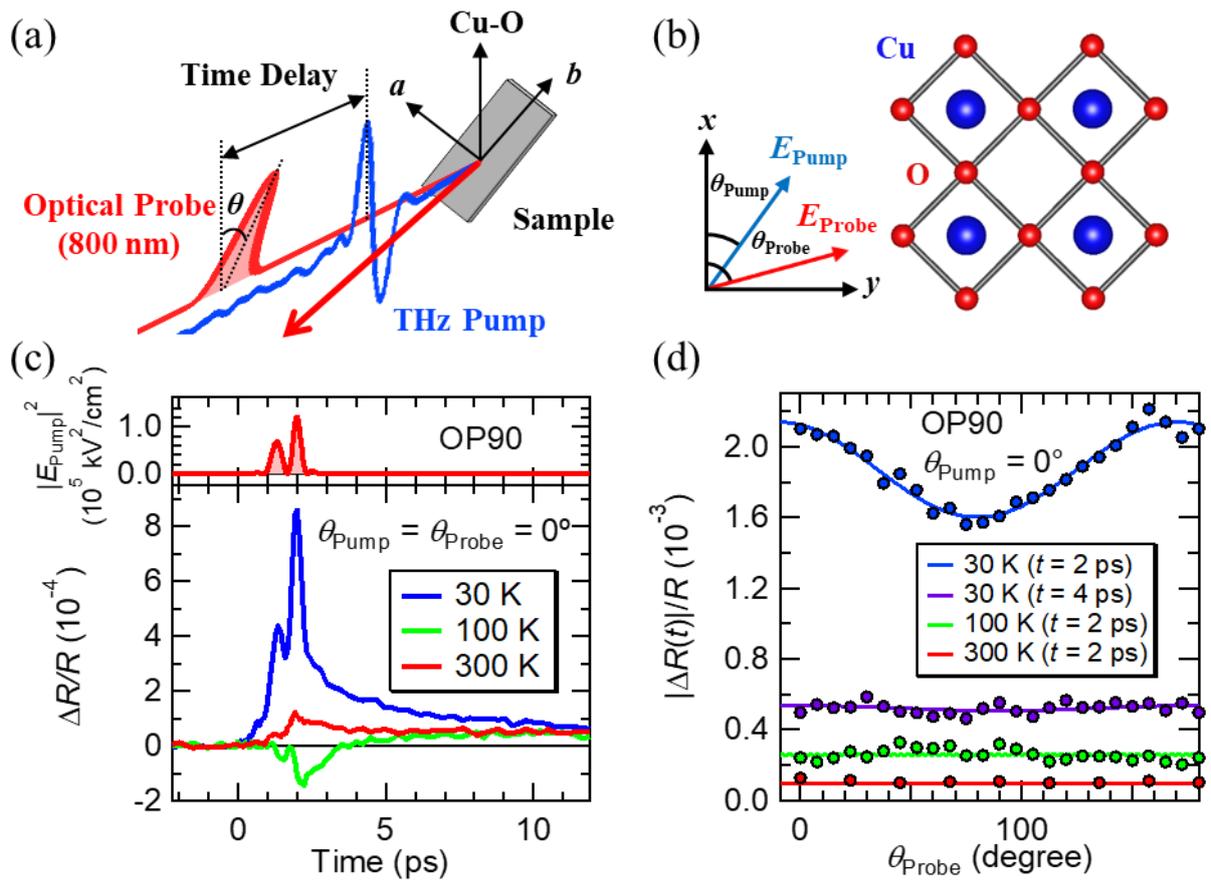

Figure 1

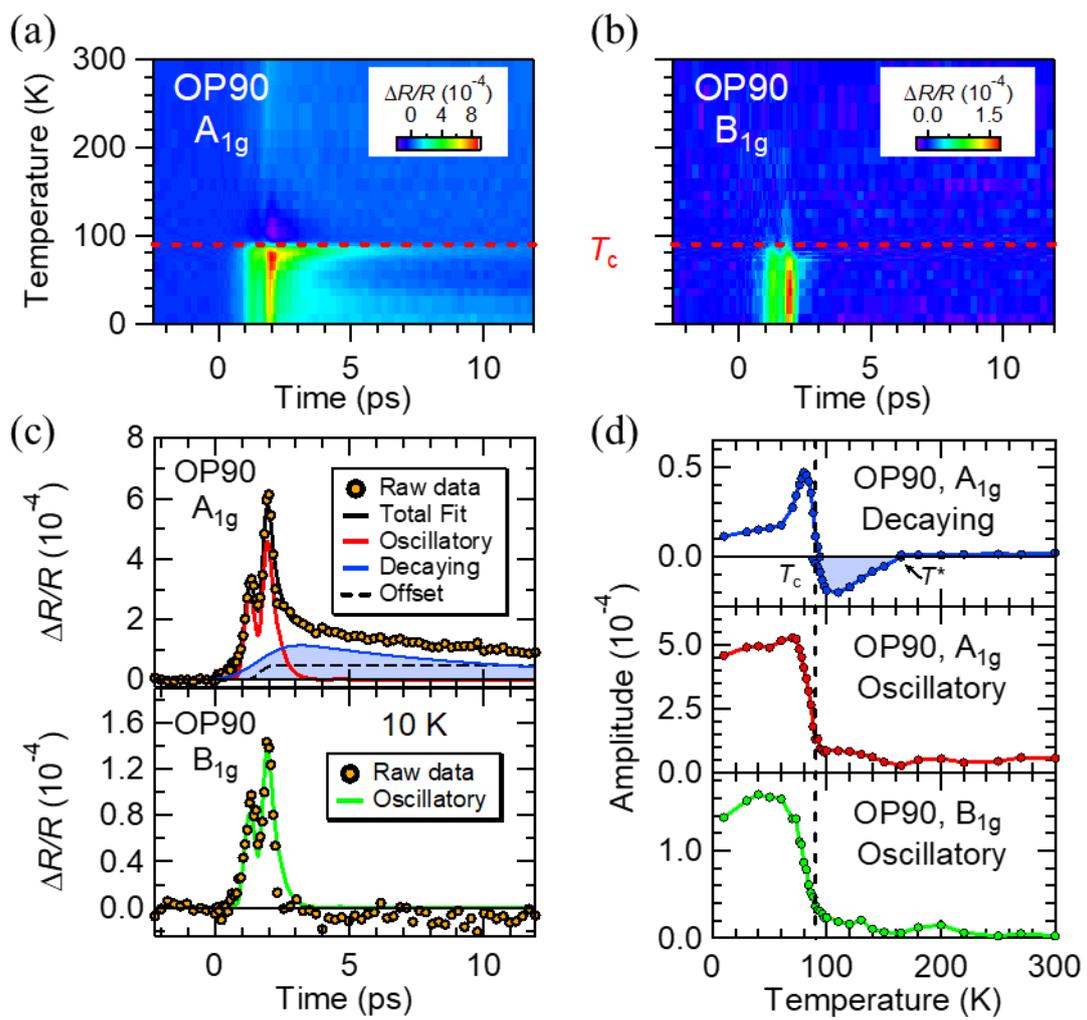

Figure 2

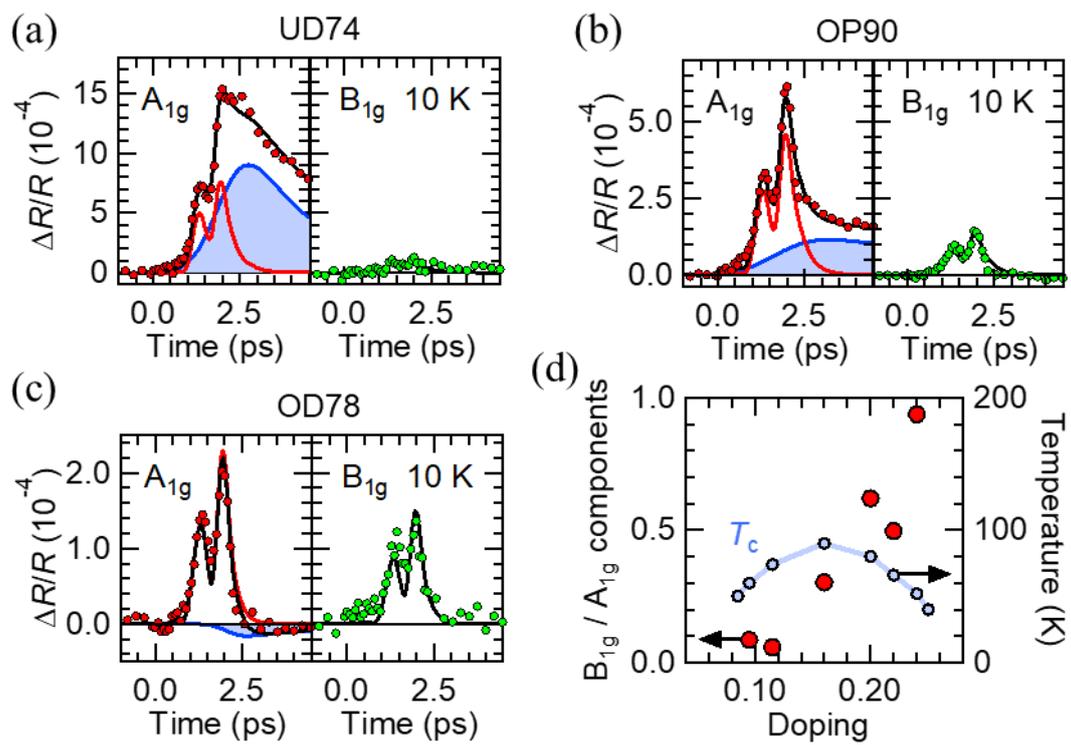

Figure 3

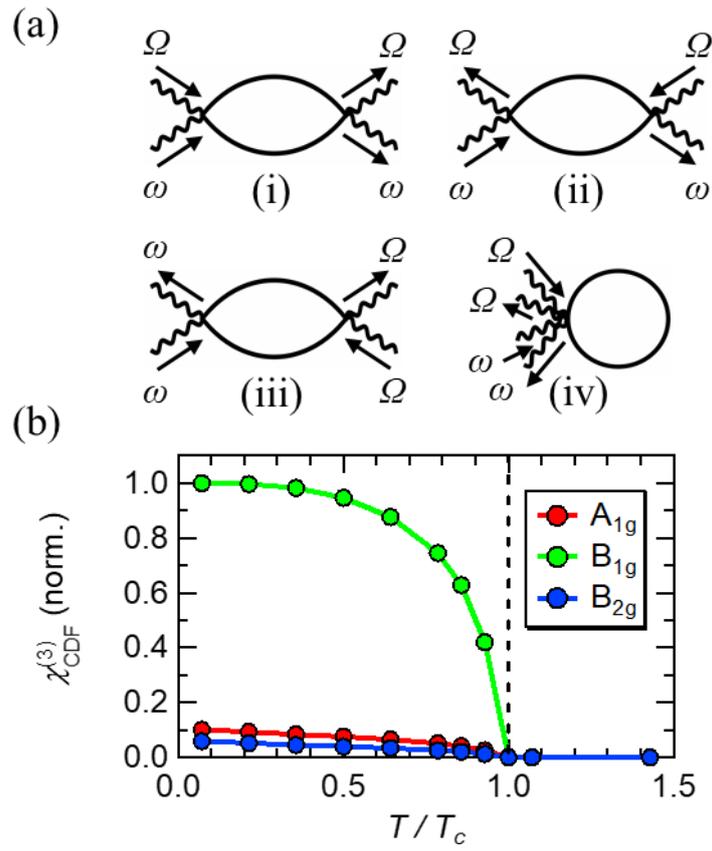

Figure 4

Table 1

|  | $A_{1g}$ | $B_{1g}$ | $B_{2g}$ |
|---|---|---|---|
| CDF | ✓ | ✓ | ✓ |
| Higgs | ✓ | 0 | 0 |

# Supplemental Material for
# "Higgs mode in the $d$-wave superconductor $Bi_2Sr_2CaCu_2O_{8+x}$ driven by an intense terahertz pulse"

Kota Katsumi, Naoto Tsuji, Yuki I. Hamada, Ryusuke Matsunaga, John Schneeloch, Ruidan D. Zhong, Genda. D. Gu, Hideo Aoki, Yann Gallais and Ryo Shimano

**Intense terahertz pulse generation**

The output from a regenerative amplified Ti:sapphire laser system with 800 nm center wavelength, 4 mJ pulse energy, 100 fs pulse duration, and 1 kHz repetition rate was divided into two beams: one for the generation of the terahertz (THz) pulse, and the other for the optical-probe pulse. To generate an intense monocycle THz pulse as an oscillating driving source, we used the tilted-pulse-front method with a $LiNbO_3$ crystal [S1] combined with the tight focusing method [S2]. The THz-pump electric field (E-field) $E_{Pump}(t)$ was detected by the electro-optic (EO) sampling in a 380 μm GaP (110) crystal placed inside the cryostat. Figure S1 shows the waveform and power spectrum of the THz-pump E-field. The peak value of the E-field reaches ~ 350 kV/cm with a central frequency ~ 0.6 THz. The near-infrared probe pulse was focused onto a 0.24 mm diameter spot on the $ab$ plane of the crystal and the THz-pump pulse was focused onto a 3.1 mm spot.

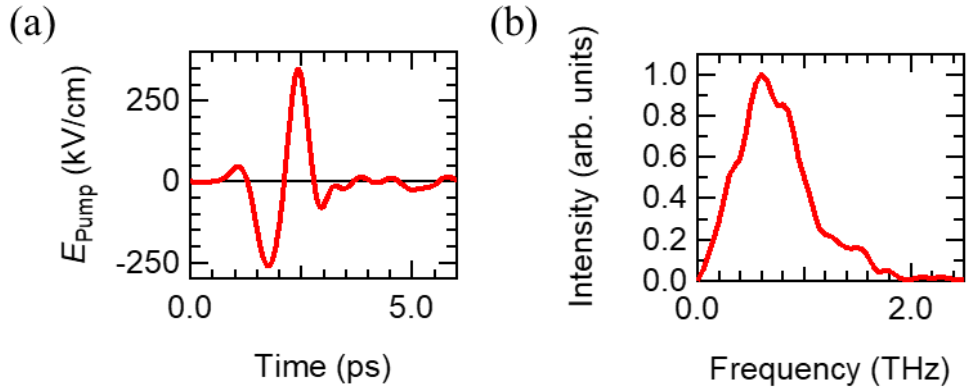

Figure S1. (a) The temporal waveform and (b) the power spectrum of the THz-pump E-field.

**THz-pump E-field dependence of the reflectivity change**

We examined the THz-pump E-field dependence of the near-infrared reflectivity change $\Delta R/R$. The THz E-field strength was continuously tuned by using three wire-grid polarizers (WGPs) inserted in the optical path of the THz pulse. Only the middle WGP was rotated to tune the THz E-field strength while keeping the waveform and the polarization identical at the sample position. Figure S2 shows the reflectivity change $\Delta R/R$ at its maximum as a function of the peak THz E-field for the OP90 sample at 10 K when $\theta_{Pump} = \theta_{Probe} = 0°$. $\Delta R_{Max}/R$ does not saturate up to ~ 350 kV/cm, indicating that the superconducting (SC) state is not significantly depleted up to the strongest THz E-field studied here.

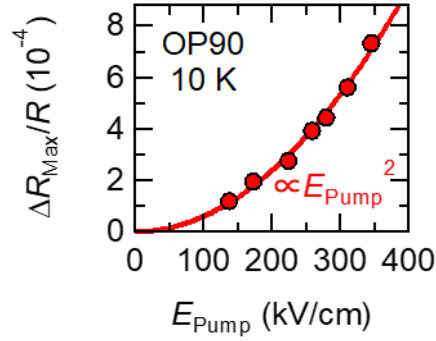

Figure S2. Dependence of the maximum amplitude of Δ$R/R$ on the peak E-field of the THz-pump pulse for OP90 at 10 K for $\theta_{Pump} = \theta_{Probe} = 0°$.

**Pump polarization dependence**

Here we show the THz-pump polarization dependence of Δ$R/R$ for the OP90 sample. We used 3 WGPs to rotate the pump polarization angle $\theta_{Pump}$ while keeping the E-field strength identical as ~ 320 kV/cm, for all the polarization angles [S3]. Figure S3(a) shows the maximum amplitude of Δ$R/R$ at 30 K when $\theta_{Probe} = 0°$ as a function of the pump polarization angle $\theta_{Pump}$. The result can be fitted by a formula $A + B \cos(2\theta_{Pump})$. This polarization-angle dependence of the pump pulse is similar to that of the probe pulse shown in Fig. 1(d). We also plot the maximum amplitude of the $A_{1g}$ and $B_{1g}$ signals against $\theta_{Pump}$ in Fig. S3(b). As expected from Eq. (2) in the main text, the $A_{1g}$ signal is angle-independent whereas the $B_{1g}$ signal follows $\cos2\theta_{Pump}$.

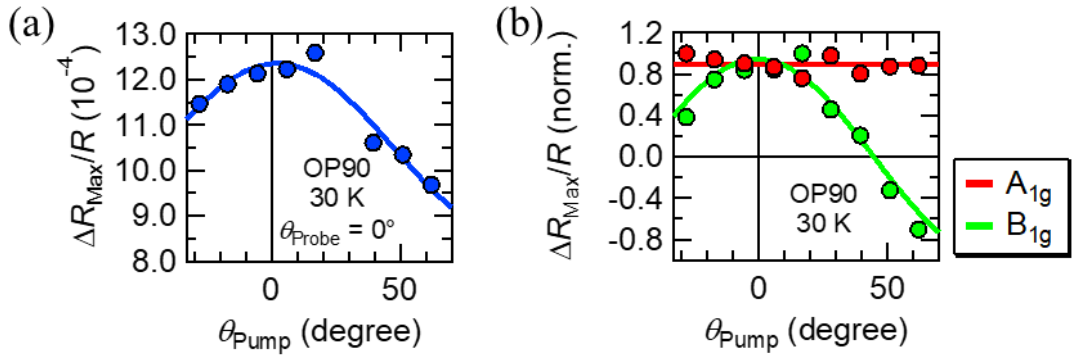

Figure S3. (a) Pump-polarization dependence of the maximum amplitude of Δ$R/R$ for $\theta_{Probe} = 0°$ at 30 K for OP90. The curve shows the fitting with a form $A + B \cos(2\theta_{Pump})$. (b) Pump-polarization dependences of the $A_{1g}$ and $B_{1g}$ Δ$R/R$ signals at 30 K. The signals are normalized by the respective maximum values. The green line is a $\cos2\theta_{Pump}$ fit for the $B_{1g}$ signal.

**The analysis for the pump and probe polarization dependence**

As we have explained in the main text, the oscillatory behavior of the observed Δ$R/R$ signal can be described by a third-order nonlinear susceptibility $\chi^{(3)}(\omega; \omega, +\Omega, -\Omega)$ [S4], where $\omega$ and $\Omega$ are the frequencies of the near-infrared pulse and THz-pump pulse, respectively. The reflectivity change Δ$R$ induced by the THz pulse can be expanded in terms of the real ($\varepsilon_1$) and imaginary ($\varepsilon_2$) parts of the dielectric constant as

$$\Delta R = \frac{\partial R}{\partial \varepsilon_1}\Delta\varepsilon_1 + \frac{\partial R}{\partial \varepsilon_2}\Delta\varepsilon_2. \tag{S1}$$

Here the change $\Delta\varepsilon$ in the complex dielectric constant is connected to $\chi^{(3)}$ as

$$\Delta\varepsilon_{ijkl} = \varepsilon_0 \chi^{(3)}_{ijkl} E^{\text{Pump}}_k E^{\text{Pump}}_l, \tag{S2}$$

where $E^{\text{Pump}}_k$ denotes the $k$th component of the THz-pump E-field while $i$ and $j$ are the indices for the probe E-field. Eq. (S2) corresponds to Eq. (1) in the main text. As we shall show, when the pump photon energy is much smaller than half the SC gap energy $\Delta_0$, the instantaneous response is off-resonant and thereby dominated by the real part of $\chi^{(3)}$. In that case we have

$$\frac{\Delta R}{R}(E^{\text{Probe}}_i, E^{\text{Probe}}_j) \sim \frac{1}{R}\frac{\partial R}{\partial \varepsilon_1}\varepsilon_0 \text{Re}\chi^{(3)}_{ijkl} E^{\text{Pump}}_k E^{\text{Pump}}_l. \tag{S3}$$

Since the experiments were performed only using the polarizations parallel to the CuO$_2$ planes, we can focus on the plane on which we can assign the axes $x$, $y$ along the Cu-O bonds (Fig. 1(b) in the main text). Here we define the pump and probe E-fields as $\boldsymbol{E}^{\text{Pump}} = E^{\text{Pump}}(\cos\theta_{\text{Pump}}, \sin\theta_{\text{Pump}})$ and $\boldsymbol{E}^{\text{Probe}} = E^{\text{Probe}}(\cos\theta_{\text{Probe}}, \sin\theta_{\text{Probe}})$. Assuming tetragonal symmetry for Bi2212, the polarization-angle dependence of $\chi^{(3)}$ can be analyzed in terms of the irreducible representations of $D_{4h}$ point group. We can then decompose $\Delta\varepsilon(\theta_{\text{Pump}}, \theta_{\text{Probe}})$ in terms of just three symmetry components of $\chi^{(3)}$ as

$$\chi^{(3)}(\theta_{\text{Pump}}, \theta_{\text{Probe}}) = \frac{1}{2}(\chi^{(3)}_{A_{1g}} + \chi^{(3)}_{B_{1g}}\cos2\theta_{\text{Pump}}\cos2\theta_{\text{Probe}} + \chi^{(3)}_{B_{2g}}\sin2\theta_{\text{Pump}}\sin2\theta_{\text{Probe}}), \tag{S4}$$

where $\chi^{(3)}_{A_{1g}} = \chi^{(3)}_{xxxx} + \chi^{(3)}_{xxyy}$, $\chi^{(3)}_{B_{1g}} = \chi^{(3)}_{xxxx} - \chi^{(3)}_{xxyy}$, and $\chi^{(3)}_{B_{2g}} = \chi^{(3)}_{xyxy} + \chi^{(3)}_{xyyx}$, and the equation corresponds to Eq. (2) in the main text. By substituting Eq. (S4) into Eq. (S3), we can express the polarization-angle dependence of $\Delta R/R$ as

$$\frac{\Delta R}{R}(\theta_{\text{Pump}}, \theta_{\text{Probe}}) = \frac{\Delta R_{A_{1g}}}{R} + \frac{\Delta R_{B_{1g}}}{R}\cos2\theta_{\text{Pump}}\cos2\theta_{\text{Probe}} + \frac{\Delta R_{B_{2g}}}{R}\sin2\theta_{\text{Pump}}\sin2\theta_{\text{Probe}}, \tag{S5}$$

where we have defined

$$\frac{\Delta R_\alpha}{R}(E^{\text{Probe}}_i, E^{\text{Probe}}_j) \sim \frac{\varepsilon_0}{2R}\frac{\partial R}{\partial \varepsilon_1}|E^{\text{Pump}}|^2 \text{Re}\,\chi^{(3)}_\alpha \quad (\alpha = A_{1g}, B_{1g}, B_{2g}). \tag{S6}$$

**The B$_{2g}$ component for $\theta_{\text{Pump}} = 45°$**

From Eq. (S6), we can obtain the B$_{1g}$ and B$_{2g}$ components of $\Delta R/R$ as

$$\frac{\Delta R_{B_{1g}}}{R} = \frac{\Delta R}{R}(\theta_{\text{Pump}} = 0°, \theta_{\text{Probe}} = 0°) - \frac{\Delta R}{R}(\theta_{\text{Pump}} = 0°, \theta_{\text{Probe}} = 90°), \tag{S7}$$

$$\frac{\Delta R_{B_{2g}}}{R} = \frac{\Delta R}{R}(\theta_{\text{Pump}} = 45°, \theta_{\text{Probe}} = 45°) - \frac{\Delta R}{R}(\theta_{\text{Pump}} = 45°, \theta_{\text{Probe}} = -45°). \tag{S8}$$

Figure S4 shows the obtained B$_{1g}$ and B$_{2g}$ components at 30 K for OP90 with the E-field strength fixed to ~ 320 kV/cm. The B$_{2g}$ signal was not resolved within the sensitivity of our measurement.

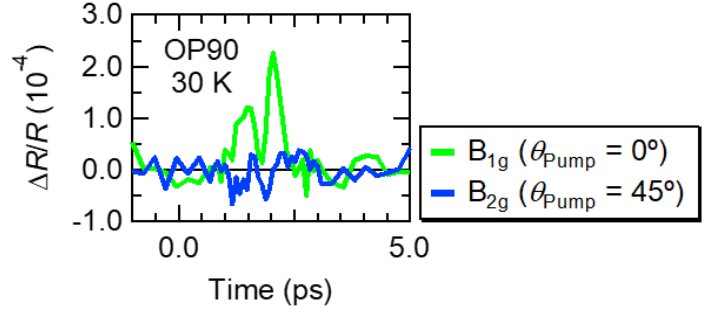

Figure S4. The $B_{1g}$ component of $\Delta R/R$ for $\theta_{Pump} = 0°$ (green), and the $B_{2g}$ component of $\Delta R/R$ for $\theta_{Pump} = 45°$ (blue) at 30 K for OP90.

**The fitting procedure**

To obtain detailed information on the temperature dependence of the symmetry-resolved components, we fitted the transient signals with a formula,

$$\frac{\Delta R(t)}{R} = A \int_{-\infty}^{\infty} |E_{Pump}(t')|^2 \exp\left(-\frac{t-t'}{\tau}\right) dt' + B \exp\left(-\frac{t-t_0}{\tau_d}\right)\left[1 - \text{erf}\left(\frac{-4(t-t_0)\tau_d + \tau_r^2}{2\sqrt{2}\tau_d \tau_r}\right)\right] \quad \text{(S9)}$$
$$+ C\left[1 - \text{erf}\left(\frac{-\sqrt{2}(t-t_0)}{\tau_r'}\right)\right].$$

The first term represents the oscillatory component or THz Kerr signal of $|E_{Pump}(t)|^2$, where $|E_{Pump}(t)|^2$ was convoluted with a single exponential function with a decay constant, $\tau \sim 200$ fs, to account for the delay in the nonlinear response of the sample [S5]. The second term corresponds to the decaying component with rise time $\tau_r$ and decay time $\tau_d$. The last term is the offset component with a rise time $\tau_r'$. Note that the $B_{1g}$ signal was fitted using only the first term ($B = C = 0$). The fitting parameters for the $A_{1g}$ signal were $A$, $B$, $C$, $t_0$ and $\tau_d$, and the other parameters were fixed to reproduce the experimental results at all temperatures. The fitting curves for the $A_{1g}$ and $B_{1g}$ signals at 10 K are shown in Fig. 2(c). The changes in $C$ are within 40 % of that at 10 K and $t_0$ varies within 1 ps.

The fitting result for $\tau_d$ below $T_c$ is shown in Fig. S5. With increasing temperature below $T_c$, $\tau_d$ first decreases, then reaches a minimum $\sim 1$ ps at 70 K, and finally shows an increase toward $T = T_c$. The behavior of $\tau_d$ far below $T_c$ can be explained by Rothwarf-Taylor (RT) model [S6]. According to the RT model, $\tau_d^{-1}$ far below $T_c$ is proportional to the number of thermally-excited quasiparticles (QPs) which increases with increasing temperature [S7,S8]. On the contrary, the increase of $\tau_d$ when $T_c$ is approached can be interpreted as the manifestation of a divergence proportional to the inverse of the SC gap energy $\Delta^{-1}$ predicted by theoretical calculations [S9]. This divergent-like behavior was also observed in previous optical pump-optical probe (OPOP) measurements for Bi2212 [S8,S10,S11]. Above $T_c$, the decay time $\tau_d$ was fixed to 0.5 ps to reproduce the experimental result. This temperature-independent decay time is consistent with that observed in OPOP measurements in the pseudogap (PG) temperature region [S8,S10]. The behavior of $\tau_d$ and the sign changes of the decaying component at $T_c$ and $T^*$ displayed in Fig. 2(d) in the main text suggest that the positive decaying component below $T_c$ and the negative decaying component above $T_c$ can be assigned to QP relaxation in the SC and PG states, respectively.

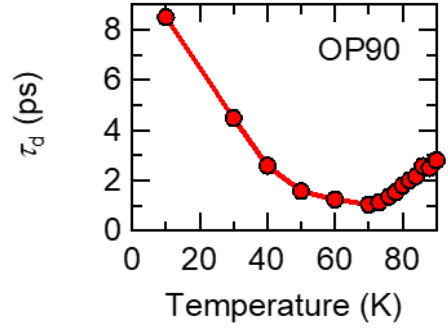

Figure S5. Temperature dependence of the decay time $\tau_d$ of the $A_{1g}$ decaying component for OP90 below $T_c$ (= 90 K).

**Temperature dependences for UD74 and OD78**

Here we show the temperature dependences of the $A_{1g}$ and $B_{1g}$ signals for the UD74 sample in Figs. S6(a) and S6(b), and for the OD78 sample in Figs. S6(c) and S6(d). In both samples, the oscillatory $A_{1g}$ and $B_{1g}$ signals sharply increase below $T_c$. The decaying component at longer delays ($t > 4$ ps) is only observed in the $A_{1g}$ symmetry. The decaying component of UD74 is positive below $T_c$ and changes its sign twice at $T_c$ and $T^*$. By contrast, the decaying component of OD78 is negative below $T_c$ and becomes positive near above $T_c$. These sign changes that depend on both temperature and doping are in good agreement with previously reported OPOP measurements in Bi2212 [S10–S12].

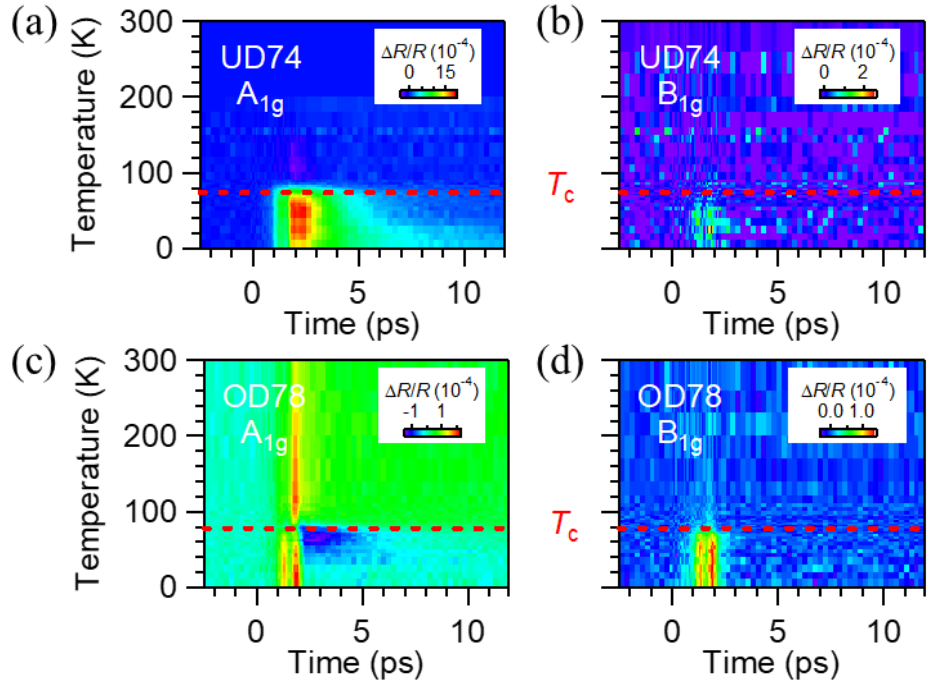

Figure S6. (a), (b) Temperature dependences of the $A_{1g}$ and $B_{1g}$ $\Delta R/R$ signals for UD74 and (c), (d) for OD78 for $\theta_{Pump} = 0°$.

**Comparison of the $A_{1g}$ and $B_{1g}$ signals for UD58, OD66 and OD52**

The pump-probe delay dependence of $A_{1g}$ and $B_{1g}$ signals for the UD58, OD66 and OD52 samples at 10 K are shown in Fig. S7. In the UD58 sample, the $A_{1g}$ oscillatory component is much larger than $B_{1g}$ component, and the $A_{1g}$ decaying component is positive. In the OD66 and OD52 samples, the $A_{1g}$ oscillatory component is comparable to the $B_{1g}$ component, and the $A_{1g}$ decaying component is negative. In all the samples studied here, the oscillatory component is always positive, while the decaying component is positive for the UD and OP samples and negative for the OD samples at 10 K.

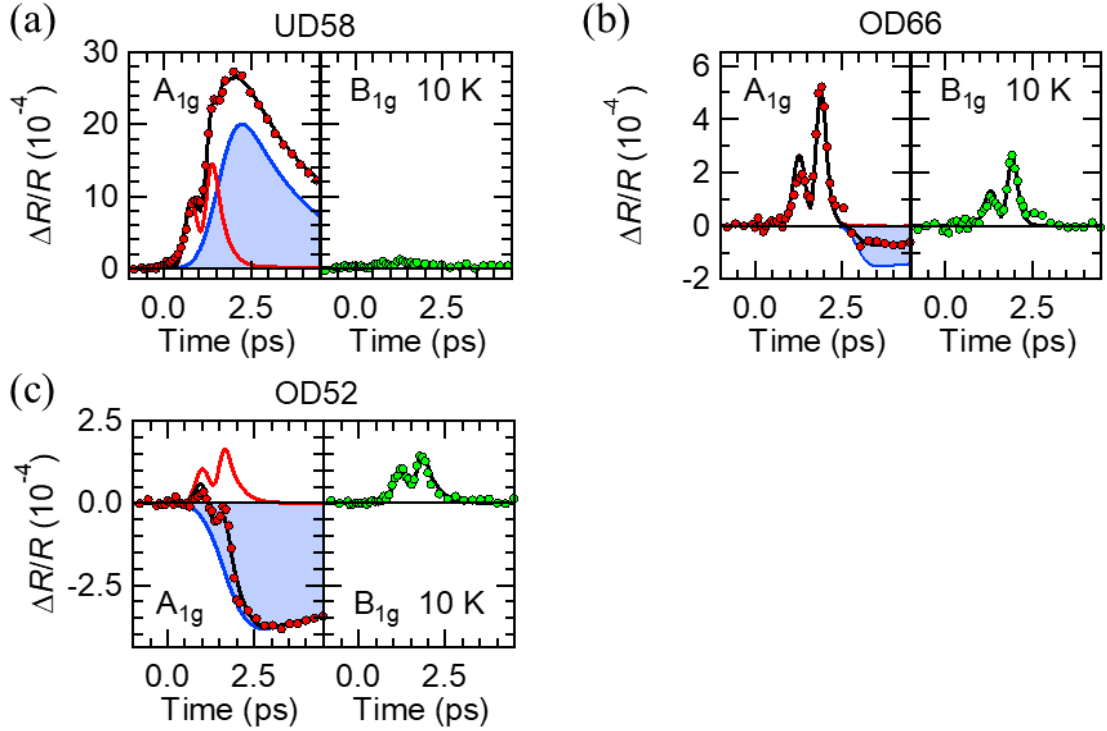

Figure S7. (a)-(c) The pump-probe delay dependence of $A_{1g}$ and $B_{1g}$ $\Delta R/R$ signals for UD58, OD66 and OD52 at 10 K for $\theta_{\text{Pump}} = 0°$. The red and blue curves show the fitting curves for the $A_{1g}$ oscillatory components and decaying components, respectively, while the black curves are the fitting curves with Eq. (S2).

**Numerical calculation of the nonlinear optical susceptibility**

The nonlinear optical susceptibility that contributes to the THz pump-optical probe (TPOP) signal is evaluated within the mean-field treatment. Here we take a pairing Hamiltonian,

$$H = \sum_{k\sigma} \xi_k c^\dagger_{k\sigma} c_{k\sigma} - \frac{1}{N} \sum_{kk'} V(k, k') c^\dagger_{k\uparrow} c^\dagger_{-k\downarrow} c_{-k'\downarrow} c_{k'\uparrow}, \quad (S10)$$

where $c^\dagger_{k\sigma}$ is the creation operator for electrons with momentum $k$ and spin $\sigma$, $\xi_k$ is the band dispersion, $N$ is the number of $k$-points, and $V(k, k')$ is the pairing interaction. We assume the $d$-wave pairing interaction of the form $V(k, k') = V u_k u_{k'}$ with $V > 0$ and $u_k = \cos k_x - \cos k_y$. We define the SC gap function,

$$\Delta_k = \frac{1}{N} \sum_{k'} V(k, k') \langle c^\dagger_{k'\uparrow} c^\dagger_{-k'\downarrow} \rangle, \quad (S11)$$

which satisfies the self-consistent mean-field gap equation,

$$\Delta_k = \frac{1}{N} \sum_{k'} V(k, k') \frac{\Delta_{k'}}{2E_{k'}} \tanh\left(\frac{E_{k'}}{2T}\right), \quad (S12)$$

where $E_k = \sqrt{\xi_k^2 + \Delta_k^2}$ is the eigenenergy of quasiparticles, and $T$ is the temperature. One can factor out the momentum dependence of the gap function as $\Delta_k = \Delta u_k$.

The dynamics of the superconductor is described by the evolution of Anderson's pseudospin $\sigma_k = \frac{1}{2} \langle \Psi_k^\dagger \boldsymbol{\tau} \Psi_k \rangle$ [S13–S15], where $\Psi_k^\dagger = (c_{k\uparrow}^\dagger, c_{-k\downarrow})$ is the Nambu spinor and $\boldsymbol{\tau} = (\tau_x, \tau_y, \tau_z)$ are the Pauli matrices. The equation of motion for the pseudospins is given by a Bloch equation,

$$\frac{\partial}{\partial t} \sigma_k(t) = 2 \boldsymbol{b}_k(t) \times \sigma_k(t). \tag{S13}$$

Here $\boldsymbol{b}_k(t) = (-\Delta_k'(t), -\Delta_k''(t), \frac{\xi_{k-A(t)} + \xi_{k+A(t)}}{2})$ is the pseudomagnetic field acting on the pseudospin, $\Delta_k'(t)$ and $\Delta_k''(t)$ are respectively the real and imaginary parts of the gap function, while $A(t) = A_\text{Pump}(t) + A_\text{Probe}(t)$ represents the vector potential for the pump and probe lasers. If we denote the deviation of the pseudospin configuration from the equilibrium state as $\sigma_k(t) = \sigma_{k,\text{eq}} + \delta\sigma_k(t)$, then $\delta\sigma_k(t)$ is even-order in $A(t)$.

The current is expressed in terms of the pseudospins as

$$\boldsymbol{j}(t) = \frac{1}{N} \sum_{k\sigma} \boldsymbol{v}_{k-A(t)} \langle c_{k\sigma}^\dagger c_{k\sigma} \rangle = \frac{1}{N} \sum_k (\boldsymbol{v}_{k-A(t)} - \boldsymbol{v}_{k+A(t)}) \left(\sigma_k^z(t) + \frac{1}{2}\right) + \text{const.} \tag{S14}$$

with $\boldsymbol{v}_k = \frac{\partial \xi_k}{\partial \boldsymbol{k}}$ the group velocity. The leading pump-probe response is third-order in $A(t)$ as

$$\boldsymbol{j}^{(3)}(t) = -\frac{2}{N} \sum_{k,i} \frac{\partial v_k}{\partial k_i} A_i(t) \delta\sigma_k^z(t) - \frac{1}{3N} \sum_{k,ijk} \frac{\partial^3 v_k}{\partial k_i \partial k_j \partial k_k} A_i(t) A_j(t) A_k(t) \left(\sigma_{k,\text{eq}}^z + \frac{1}{2}\right). \tag{S15}$$

Let us assume a sinusoidal form for the pump and probe E-fields,

$$A_\text{Pump}(t) = 2 A_\text{Pump} \cos \Omega t = A_\text{Pump} (e^{i\Omega t} + e^{-i\Omega t}), \tag{S16}$$

$$A_\text{Probe}(t) = A_\text{Probe} e^{-i\omega t}, \tag{S17}$$

where $\Omega$ and $\omega$ are the frequencies of the pump and probe light, respectively. We define the polarization vectors $\boldsymbol{e}^\text{Pump}$ and $\boldsymbol{e}^\text{Probe}$ for the pump and probe light as $A_\text{Pump} = \boldsymbol{e}^\text{Pump} A_\text{Pump}$ and $A_\text{Probe} = \boldsymbol{e}^\text{Probe} A_\text{Probe}$ ($|\boldsymbol{e}^\text{Pump}| = |\boldsymbol{e}^\text{Probe}| = 1$). The nonlinear current $\boldsymbol{j}^{(3)}(t)$ that contributes to the pump-probe spectroscopy has the same time dependence as that of the probe light ($\propto e^{-i\omega t}$). Hence $\boldsymbol{j}^{(3)}(t)$ must contain the product of $A_\text{Pump} e^{i\Omega t}$, $A_\text{Pump} e^{-i\Omega t}$, and $A_\text{Probe} e^{-i\omega t}$. The third-order nonlinear optical susceptibility $\chi^{(3)}$ that represents the pump-probe signal is defined by

$$\boldsymbol{e}_\text{Probe} \cdot \boldsymbol{j}^{(3)}(t) = \chi^{(3)}(\omega; \omega, +\Omega, -\Omega) A_\text{Pump}^2 A_\text{Probe} e^{-i\omega t}. \tag{S18}$$

For the first term in Eq. (S15), there are three possibilities:

(1) $A_i(t)$ is $A_\text{Pump} e^{i\Omega t}$ with $\delta\sigma_k^z(t)$ containing $A_\text{Pump} e^{-i\Omega t}$ and $A_\text{Probe} e^{-i\omega t}$.

(2) $A_i(t)$ is $A_\text{Pump} e^{-i\Omega t}$ with $\delta\sigma_k^z(t)$ containing $A_\text{Pump} e^{i\Omega t}$ and $A_\text{Probe} e^{-i\omega t}$.

(3) $A_i(t)$ is $A_\text{Probe} e^{-i\omega t}$ with $\delta\sigma_k^z(t)$ containing $A_\text{Pump} e^{i\Omega t}$ and $A_\text{Pump} e^{-i\Omega t}$.

The second term in Eq. (S15), on the other hand, has a unique possibility that $A_i(t)$, $A_j(t)$, and $A_k(t)$ are a permutation of $A_{\text{Pump}} e^{i\Omega t}$, $A_{\text{Pump}} e^{-i\Omega t}$, and $A_{\text{Probe}} e^{-i\omega t}$. Correspondingly, we have four different diagrams for $\chi^{(3)}$ as displayed in Fig. 4(a) in the main text.

In the case of the TPOP spectroscopy, the frequency $\omega$ of the probe light exceeds all the other relevant energy scales. In this situation, as we have discussed in the main text, the dominant contribution of the charge density fluctuation (CDF) to $\chi^{(3)}$ comes from the case (3) in the above [which corresponds to diagram (iii) in Fig. 4(a)]. The CDF contribution including the screening effect is explicitly calculated as

$$\chi^{(3)}_{\text{CDF}} = \frac{1}{N} \sum_{\mathbf{k}, ijkl} \frac{\partial^2 \xi_{\mathbf{k}}}{\partial k_i \partial k_j} e_i^{\text{Pump}} e_j^{\text{Pump}} \frac{\partial^2 \xi_{\mathbf{k}}}{\partial k_k \partial k_l} e_k^{\text{Probe}} e_l^{\text{Probe}} \chi_{33}(\mathbf{k}, 0)$$
$$- \frac{\left[\frac{1}{N} \sum_{\mathbf{k}, ij} \frac{\partial^2 \xi_{\mathbf{k}}}{\partial k_i \partial k_j} e_i^{\text{Pump}} e_j^{\text{Pump}} \chi_{33}(\mathbf{k}, 0)\right] \left[\frac{1}{N} \sum_{\mathbf{k}, kl} \frac{\partial^2 \xi_{\mathbf{k}}}{\partial k_k \partial k_l} e_k^{\text{Probe}} e_l^{\text{Probe}} \chi_{33}(\mathbf{k}, 0)\right]}{\frac{1}{N} \sum_{\mathbf{k}} \chi_{33}(\mathbf{k}, 0)}, \quad \text{(S19)}$$

where $\chi_{33}(\mathbf{k}, \nu)$ is the dynamical charge susceptibility [S3,S16]. Within the mean-field theory, $\chi_{33}(\mathbf{k}, \nu)$ is given by

$$\chi_{33}(\mathbf{k}, \nu) = \frac{2\Delta_{\mathbf{k}}^2}{E_{\mathbf{k}}[4E_{\mathbf{k}}^2 - (\nu + i\delta)^2]} \tanh\left(\frac{E_{\mathbf{k}}}{2T}\right) \quad \text{(S20)}$$

with $\delta$ an infinitesimal positive constant (in practice we take a finite value for $\delta$ to regularize the divergence in $\chi_{33}$).

To investigate the polarization-angle dependence of $\chi^{(3)}$, we set $\mathbf{e}^{\text{Pump}} = (\cos\theta_{\text{Pump}}, \sin\theta_{\text{Pump}}, 0)$ and $\mathbf{e}^{\text{Probe}} = (\cos\theta_{\text{Probe}}, \sin\theta_{\text{Probe}}, 0)$. Assuming tetragonal symmetry for Bi2212, we can decompose the nonlinear susceptibility $\chi^{(3)}_{\text{CDF}}$ into the irreducible representations of $D_{4h}$ point group as in Eq. (2) in the main text. These components are given as

$$\chi^{(3)}_{\text{CDF}, A_{1g}} = \frac{1}{N} \sum_{\mathbf{k}} \left(\frac{\partial^2 \xi_{\mathbf{k}}}{\partial k_x^2}\right)^2 \chi_{33}(\mathbf{k}, 0) + \frac{1}{N} \sum_{\mathbf{k}} \frac{\partial^2 \xi_{\mathbf{k}}}{\partial k_x^2} \frac{\partial^2 \xi_{\mathbf{k}}}{\partial k_y^2} \chi_{33}(\mathbf{k}, 0) - 2 \frac{\left[\frac{1}{N} \sum_{\mathbf{k}} \frac{\partial^2 \xi_{\mathbf{k}}}{\partial k_x^2} \chi_{33}(\mathbf{k}, 0)\right]^2}{\frac{1}{N} \sum_{\mathbf{k}} \chi_{33}(\mathbf{k}, 0)}, \quad \text{(S21)}$$

$$\chi^{(3)}_{\text{CDF}, B_{1g}} = \frac{1}{N} \sum_{\mathbf{k}} \left(\frac{\partial^2 \xi_{\mathbf{k}}}{\partial k_x^2}\right)^2 \chi_{33}(\mathbf{k}, 0) - \frac{1}{N} \sum_{\mathbf{k}} \frac{\partial^2 \xi_{\mathbf{k}}}{\partial k_x^2} \frac{\partial^2 \xi_{\mathbf{k}}}{\partial k_y^2} \chi_{33}(\mathbf{k}, 0), \quad \text{(S22)}$$

$$\chi^{(3)}_{\text{CDF}, B_{2g}} = \frac{2}{N} \sum_{\mathbf{k}} \left(\frac{\partial^2 \xi_{\mathbf{k}}}{\partial k_x \partial k_y}\right)^2 \chi_{33}(\mathbf{k}, 0). \quad \text{(S23)}$$

The Higgs-mode contribution to $\chi^{(3)}$ is also classified in the same way as CDF. The relevant diagrams are those corresponding to (i)-(iii) in Fig. 4(a) in the main text with the vertex function inserted inside the bubbles. The dominant contribution for the TPOP spectroscopy comes from the frequency-independent one that corresponds to diagram (iii). For this type of the diagram, the Higgs-mode contribution appears only in the $A_{1g}$ component, i.e.,

$$\chi^{(3)}_{\text{Higgs}}(\theta_{\text{Pump}}, \theta_{\text{Probe}}) = \frac{1}{2}\chi^{(3)}_{\text{Higgs, A}_{1g}}. \tag{S24}$$

This sharply contrasts with the polarization dependence of the CDF contribution (see Table 1 in the main text), which allows us to discriminate the CDF and Higgs-mode contributions in TPOP spectroscopy experiments. Within the mean-field theory, the $A_{1g}$ component of the Higgs-mode contribution (including the screening effect) is explicitly evaluated as

$$\chi^{(3)}_{\text{Higgs, A}_{1g}} = 2V\left[1 - V\left(\frac{1}{N}\sum_k u_k^2\, \chi_{11}(k,0) - \frac{\left(\frac{1}{N}\sum_k u_k \chi_{31}(k,0)\right)^2}{\frac{1}{N}\sum_k \chi_{33}(k,0)}\right)\right]^{-1}$$
$$\times \left(\frac{1}{N}\sum_k \frac{\partial^2 \xi_k}{\partial k_x^2} u_k\, \chi_{31}(k,0) - \frac{\frac{1}{N}\sum_k \frac{\partial^2 \xi_k}{\partial k_x^2}\chi_{33}(k,0)\frac{1}{N}\sum_k u_k\, \chi_{31}(k,0)}{\frac{1}{N}\sum_k \chi_{33}(k,0)}\right)^2. \tag{S25}$$

Here $\chi_{11}(k, \nu)$ and $\chi_{31}(k, \nu)$ are amplitude-amplitude and amplitude-charge dynamical susceptibilities, respectively. In the mean-field theory, they are calculated as

$$\chi_{11}(k, \nu) = \frac{2\xi_k^2}{E_k[4E_k^2 - (\nu + i\delta)^2]}\tanh\left(\frac{E_k}{2T}\right), \tag{S26}$$

$$\chi_{31}(k, \nu) = -\frac{2\xi_k \Delta_k}{E_k[4E_k^2 - (\nu + i\delta)^2]}\tanh\left(\frac{E_k}{2T}\right). \tag{S27}$$

To numerically evaluate these quantities for Bi2212, we employ a single-band tight-binding model with the band dispersion,

$$\xi_k = -2t(\cos k_x + \cos k_y) + 4t'\cos k_x \cos k_y - 2t''(\cos 2k_x + \cos 2k_y) - \mu, \tag{S28}$$

where $t$, $t'$ and $t''$ are respectively the nearest, second-, and third-neighbor hoppings on the two-dimensional square lattice, and $\mu$ is the chemical potential. We adopt $t'/t=0.2$ and $t''/t = 0.1$ from the literature [S17], and take $V/t = 1$ and $\delta/t = 0.01$. The filling is set to be 20 % hole doped. The result for the CDF contribution is shown in Fig. 4(b) in the main text, while that for the Higgs-mode contribution is shown in Fig. S8 here. One can see that the $A_{1g}$ component grows below $T_c$, evidencing the correlation with superconductivity. Note that $\chi^{(3)}_{\text{Higgs}}$ in Fig. S6 is normalized by its maximum value for the lowest temperature considered, while $\chi^{(3)}_{\text{CDF}}$ in Fig. 4(b) is normalized by its maximum value of the $B_{1g}$ component for the lowest temperature considered, which is 80 times larger than that of $\chi^{(3)}_{\text{Higgs}}$. In general, the magnitude of $\chi^{(3)}_{\text{Higgs}}$ in the mean-field treatment is much smaller than that of $\chi^{(3)}_{\text{CDF}}$. This situation is similar to the $s$-wave case: The Higgs-mode contribution for the susceptibility of the THG is suppressed by a factor of $(\Delta/V)^2$ as compared to the CDF contribution [S18]. However, this is just an artifact of the mean-field approximation [S16]. For instance, if one takes account of strong correlation effects such as the retarded phonon-mediated interaction in the $s$-wave case, the Higgs-mode contribution can be comparable to, or even larger than, the CDF contribution. In contrast to the relative magnitudes, we can note that the polarization-angle dependence of the CDF or Higgs-mode contribution remains almost unchanged when one goes beyond the mean-field theory. It is natural to expect a similar behavior for the case of $d$-wave superconductors.

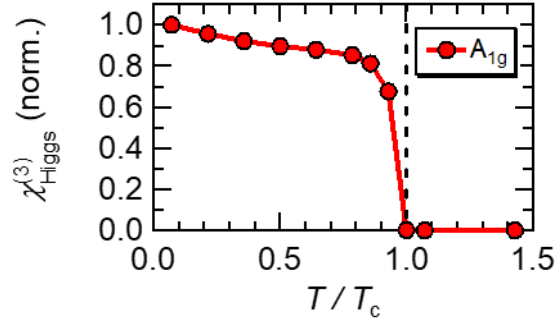

Figure S8. The mean-field result for the Higgs-mode contribution to the nonlinear optical susceptibility $\chi^{(3)}$ for the TPOP spectroscopy. $\chi^{(3)}_{\text{Higgs}}$ is here normalized by its maximum value for the lowest $T$ considered.

**Dominance of the $B_{1g}$ component in the CDF contribution**

As we have seen in Fig. 4(b) in the main text, the CDF contribution to the nonlinear optical susceptibility for the TPOP spectroscopy is dominated by the $B_{1g}$ component. This can be understood from the microscopic expression for $\chi^{(3)}_{\text{CDF}}$ in Eqs. (S21)-(S23). The $B_{2g}$ component is smaller than the other components because the cross derivative $\left(\frac{\partial^2 \xi_{\boldsymbol{k}}}{\partial k_x \partial k_y}\right)^2$ in Eq. (S23) is proportional to $(t')^2$ ($\ll t^2$), while the other factors $\left(\frac{\partial^2 \xi_{\boldsymbol{k}}}{\partial k_x^2}\right)^2$ and $\frac{\partial^2 \xi_{\boldsymbol{k}}}{\partial k_x^2}\frac{\partial^2 \xi_{\boldsymbol{k}}}{\partial k_y^2}$ in Eqs. (S21) and (S22) are basically proportional to $t^2$. To compare the magnitudes between the $A_{1g}$ and $B_{1g}$ components, let us plot the factors $\left(\frac{\partial^2 \xi_{\boldsymbol{k}}}{\partial k_x^2}\right)^2$ and $\frac{\partial^2 \xi_{\boldsymbol{k}}}{\partial k_x^2}\frac{\partial^2 \xi_{\boldsymbol{k}}}{\partial k_y^2}$ in Eqs. (S21) and (S22) in Fig. S9. The charge susceptibility $\chi_{33}(\boldsymbol{k}, 0)$ has a contribution concentrated around the Fermi surface, which is highlighted by yellow lines in Fig. S9 for the half-filled case. Near the Fermi surface, $\left(\frac{\partial^2 \xi_{\boldsymbol{k}}}{\partial k_x^2}\right)^2$ is positive (in Fig. S9(a)), while $\frac{\partial^2 \xi_{\boldsymbol{k}}}{\partial k_x^2}\frac{\partial^2 \xi_{\boldsymbol{k}}}{\partial k_y^2}$ is negative (Fig. S9(b)). Hence the first and second terms in Eq. (S21) tend to cancel with each other, whereas the first and second terms in Eq. (S22) add up. As a result, the $A_{1g}$ component is suppressed while the $B_{1g}$ component is enhanced. The screening effect (which corresponds to the third term in Eq. (S21)) further suppresses the $A_{1g}$ component.

The argument above is valid within the mean-field theory. However, we speculate that the situation is qualitatively similar in strongly correlated systems. If one takes account of strong correlation effects, $\chi_{33}(\boldsymbol{k}, 0)$ in Eq. (S19) is replaced with the one calculated beyond the mean-field theory. Due to the self-energy correction, the peaks in $\chi_{33}(\boldsymbol{k}, 0)$ spread to some extent. In the underdoped regime, the self-energy effect should be significant in the anti-nodal regions ($\boldsymbol{k} \sim (\pm\pi, 0), (0, \pm\pi)$). Therefore, the cancellation between the first and second terms in Eq. (S21) becomes less effective but remains. This will make the magnitude of the $A_{1g}$ component smaller than the $B_{1g}$ component.

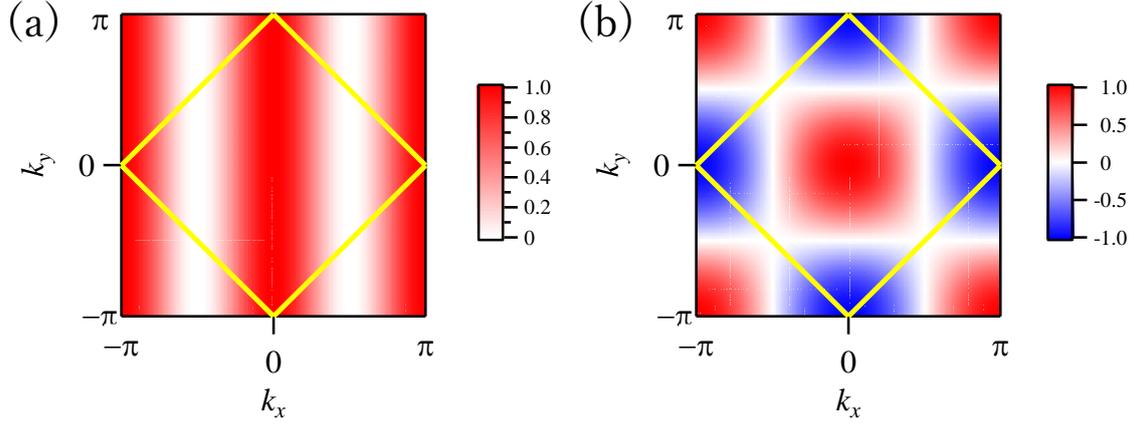

Figure S9. Color-coded $\left(\frac{\partial^2 \xi_{\bm{k}}}{\partial k_x^2}\right)^2 \sim 4t^2 \cos^2 k_x$ (a) and $\frac{\partial^2 \xi_{\bm{k}}}{\partial k_x^2}\frac{\partial^2 \xi_{\bm{k}}}{\partial k_y^2} \sim 4t^2 \cos k_x \cos k_y$ (b) in the unit of $4t^2$. Here we neglect the subleading terms involving $t'$ and $t''$. Yellow lines indicate the Fermi surface at half filling.